# Quantum Electrodynamics with Anharmonic Waves

F. J. Himpsel

Physics Department, University of Wisconsin Madison, Madison WI 53706, USA

**Abstract**

This is the second step of a program to use anharmonic plane waves as basis set in non-perturbative quantum field theory. The general framework developed previously is applied to quantum electrodynamics. To test the compatibility with standard quantum electrodynamics, the Feynman rules are generalized to anharmonic waves by expanding the field operators into anharmonic plane waves. A sum rule for the Fourier coefficients of anharmonic waves ensures that the generalized Feynman rules are equivalent to the standard rules after summing over all harmonics. It is possible to construct diagrams for the generation of harmonics. They are of $O(\alpha^2)$ for photons and of $O(\alpha)$ for electrons. To tackle intrinsically non-perturbative phenomena it is proposed to insert anharmonic field operators into the Dyson-Schwinger equations while retaining only the lowest harmonics.

―――――――――――――――――――――――
Electronic address: fhimpsel@wisc.edu

**Contents**

## 1. Anharmonic Vector and Spinor Fields in Quantum Electrodynamics

There has been long-standing interest in developing quantum electrodynamics beyond perturbation theory for nonlinear quantum effects at high fields. The nonlinear Heisenberg-Euler Lagrangian has become a widespread model [1],[2]. Rather than altering the Lagrangian of quantum electrodynamics, this work aims at non-perturbative solutions of the standard Lagrangian. Exact field operators are expanded into anharmonic plane waves [3]. The topic of interest is the internal stability of the electron, another non-perturbative problem with a long track record. It implies the creation of a new bound state, while perturbation theory covers changes of known bound states.

A previous publication explored the most general class of plane wave solutions that fulfill the basic requirements of quantum field theory. These were found to be anharmonic generalizations of $\exp(iz)$ which satisfy three simple criteria [3]:

(1) $\quad \psi(z+2\pi) = \psi(z) \qquad\qquad z = -p_\mu x^\mu = (\mathbf{p}\,\mathbf{x} - E\,t)$

(2) $\quad \psi(z)^* \cdot \psi(z) = 1$

(3) $\quad \psi(-z)^* = \psi(z) \qquad\qquad \psi(0) = 1 \quad \psi(\pi) = -1$

The covariant variable z describes plane waves and ensures relativistic invariance. Three symmetry classes of anharmonic waves were distinguished by symmetry properties of their Fourier coefficients. Here we use Class 2, which is applicable to interactions without spontaneous symmetry breaking, such as quantum electrodynamics. The Fourier series of a Class 2 anharmonic wave has the form:

(4) $\quad \psi(z) = \sum_{n=-\infty}^{\infty} a_n \cdot \exp[i(2n+1)z] \qquad a_n \text{ real}$

The $a_n$ have to fulfill an infinite number of constraints to satisfy condition (2):

(5a) $\quad \sum_{n=-\infty}^{\infty} a_n^2 = 1 \qquad (j=0)$

(5b) $\quad \sum_{n=-\infty}^{\infty} a_n\, a_{n+j} = 0 \qquad \text{for} \quad |j| = 1,\ldots,\infty$

Since (5a,b) comprise all possible products of two Fourier coefficients, one can add up all the equations from $j=-\infty$ to $j=\infty$ to obtain the sum rule:

(5c) $\quad \sum_{n,m=-\infty}^{\infty} a_n\, a_m = 1$

In the lowest non-trivial approximation the anharmonicity is determined by a single parameter $|a_1|$:

(6a) $\quad \boxed{a_0 \approx 1 \quad\quad a_{-1} \approx -a_1}$

The approximate anti-symmetry between the lowest positive and negative harmonics can be generalized by postulating a similar requirement for all other coefficients:

(6b) $\quad a_{-n} = (-)^n \cdot a_n \quad\quad$ for $\quad n = -\infty, \ldots, +\infty$

Quantum field operators were expanded into anharmonic waves in [3] for a real scalar field. The anharmonic vector field $A_\mu$ of the photon has a similar expansion, consisting of a Class 2 Fourier series over the harmonics $(2n+1) \cdot k_\mu$. In addition, one needs to include the usual polarization vectors $\varepsilon_\mu(\mathbf{k},\lambda)$ and set the mass to zero:

$$\boxed{\begin{array}{l}(7) \quad A_\mu(x) = (2\pi)^{-3/2} \sum_{\lambda=1,2} \int d^3k \, (2E)^{-1/2} \cdot \\ \cdot \sum_{n=-\infty}^{\infty} a_n |2n+1|^{5/2} \cdot \left[ a((2n+1)\mathbf{k},\lambda) \cdot \varepsilon_\mu(\mathbf{k},\lambda) \cdot e^{-i(2n+1)kx} + a^\dagger((2n+1)\mathbf{k},\lambda) \cdot \varepsilon_\mu^*(\mathbf{k},\lambda) \cdot e^{+i(2n+1)kx} \right]\end{array}}$$

$\quad$ a = photon annihilation operator $\quad\quad$ $a^\dagger$ = photon creation operator

$\quad [a((2n+1)\mathbf{k},\lambda), a^\dagger((2n'+1)\mathbf{k}',\lambda')] = \delta_{\lambda,\lambda'} \, \delta^3((2n+1)\mathbf{k} - (2n'+1)\mathbf{k}')$

$\quad [a((2n+1)\mathbf{k},\lambda), a((2n'+1)\mathbf{k}',\lambda')] = [a^\dagger((2n+1)\mathbf{k},\lambda), a^\dagger((2n'+1)\mathbf{k}',\lambda')] = 0$

$\quad x = x^\mu = (t,\mathbf{x}) \quad k = k^\mu = (k^0,\mathbf{k}) \quad E = |k^0| = |\mathbf{k}| \quad k_\mu k^\mu = 0 \quad kx = k_\mu x^\mu$

The expansion of the anharmonic Dirac spinor field $\Psi$ is more involved, since particles and antiparticles need to be distinguished (see Appendix A). There are two sets of creation and annihilation operators ($b^\dagger, b$ for electrons and $d^\dagger, d$ for positrons). The polarization vector $\varepsilon_\mu$ is replaced by the spinors $u$ and $\hat{u}$ for electrons with positive and negative energy ($v, \hat{v}$ for positrons). These are given in (A2),(A2′). The sum over the polarizations $\lambda$ becomes the sum over the two spin orientations $s=\uparrow,\downarrow$. An extra factor $(2m_e)^{1/2}$ is needed to obtain the correct dimension of $(\text{mass})^{3/2}$ for a spinor field, in contrast to $(\text{mass})^1$ for scalar and vector fields. That generates an extra factor $|2m+1|^{1/2}$. Finally, a fermion field requires anti-commutators instead of commutators:

$$\boxed{\begin{array}{l}(8) \quad \Psi(x) = (2\pi)^{-3/2} \sum_{s=\uparrow,\downarrow} \int d^3p \, (m_e/E)^{1/2} \cdot \\ \cdot \left[ \begin{array}{l} \sum_{m\geq 0} b_m |2m+1|^3 \cdot \left[ b((2m+1)\mathbf{p},s) \cdot u(\mathbf{p},s) \cdot e^{-i(2m+1)px} + d^\dagger((2m+1)\mathbf{p},s) \cdot v(\mathbf{p},s) \cdot e^{+i(2m+1)px} \right] \\ + \sum_{m<0} b_m |2m+1|^3 \cdot \left[ \hat{b}((2m+1)\mathbf{p},s) \cdot \hat{u}(\mathbf{p},s) \cdot e^{-i(2m+1)px} + \hat{d}^\dagger((2m+1)\mathbf{p},s) \cdot \hat{v}(\mathbf{p},s) \cdot e^{+i(2m+1)px} \right] \end{array} \right]\end{array}}$$

$$\overline{\Psi}(x) = (2\pi)^{-3/2} \sum_{s=\uparrow,\downarrow} \int d^3p \, (m_e/E)^{1/2} \cdot$$

$$\cdot \left[ \begin{array}{l} \sum_{m \geq 0} b_m |2m+1|^3 \cdot \left[ b^\dagger((2m+1)\mathbf{p},s) \cdot \overline{u}(\mathbf{p},s) \cdot e^{+i(2m+1)px} + d((2m+1)\mathbf{p},s) \cdot \overline{v}(\mathbf{p},s) \cdot e^{-i(2m+1)px} \right] \\ + \sum_{m<0} b_m |2m+1|^3 \cdot \left[ \hat{b}^\dagger((2m+1)\mathbf{p},s) \cdot \overline{\hat{u}}(\mathbf{p},s) \cdot e^{+i(2m+1)px} + \hat{d}((2m+1)\mathbf{p},s) \cdot \overline{\hat{v}}(\mathbf{p},s) \cdot e^{-i(2m+1)px} \right] \end{array} \right]$$

b = electron annihilation operator $\quad$ $b^\dagger$ = electron creation operator

d = positron annihilation operator $\quad$ $d^\dagger$ = positron creation operator

$\{ b((2m+1)\mathbf{p},s), b^\dagger((2m'+1)\mathbf{p}',s') \} = \delta_{s,s'} \, \delta^3((2m+1)\mathbf{p} - (2m'+1)\mathbf{p}')$

$\{ d((2m+1)\mathbf{p},s), d^\dagger((2m'+1)\mathbf{p}',s') \} = \delta_{s,s'} \, \delta^3((2m+1)\mathbf{p} - (2m'+1)\mathbf{p}')$

Anticommutators of the type $\{b,b\}$ and $\{b^\dagger,b^\dagger\}$ vanish.

$p = p^\mu = (p^0,\mathbf{p}) \quad E = |p^0| = (\mathbf{p}^2 + m_e^2)^{1/2} \quad p_\mu p^\mu = m_e^2 \quad px = p_\mu x^\mu$

There is an alternative way to handle negative harmonics which avoids negative energy states. The standard Feynman rules suggest converting particles with negative energy into antiparticles with opposite four-momentum and spin (using CPT symmetry). For the Dirac field an incoming electron with negative energy is transformed into an outgoing positron with positive energy and vice versa. This implies the following substitutions:

$$\left. \begin{array}{l} \hat{b}((2m+1)\mathbf{p},s) \cdot \hat{u}(\mathbf{p},s) \cdot e^{-i(2m+1)px} \implies d^\dagger(|2m+1|\mathbf{p},-s) \cdot v(\mathbf{p},-s) \cdot e^{+i|2m+1|px} \\ \hat{d}^\dagger((2m+1)\mathbf{p},s) \cdot \hat{v}(\mathbf{p},s) \cdot e^{+i(2m+1)px} \implies b(|2m+1|\mathbf{p},-s) \cdot u(\mathbf{p},-s) \cdot e^{-i|2m+1|px} \end{array} \right\} \text{ for } m<0$$

The momentum $\mathbf{p}$ is not inverted for $\hat{u}$ and $\hat{v}$, because they are defined in (A2′) such that $\mathbf{p}$ is the momentum of the fundamental, while the momentum eigenvalue is $-\mathbf{p}$. This allows all terms of an anharmonic Fourier series to have a common label $p_\mu$. With these substitutions one can combine positive and negative harmonics in (8) to define the alternative field operators $\Psi'(x)$ and $\overline{\Psi}'(x)$:

(8′)
$$\Psi'(x) = (2\pi)^{-3/2} \sum_{s=\uparrow,\downarrow} \int d^3p \, (m_e/E)^{1/2} \cdot$$
$$\sum_{m=-\infty}^{\infty} b_m |2m+1|^3 \cdot \left[ b(|2m+1|\mathbf{p},s) \cdot u(\mathbf{p},s) \cdot e^{-i|2m+1|px} + d^\dagger(|2m+1|\mathbf{p},s) \cdot v(\mathbf{p},s) \cdot e^{+i|2m+1|px} \right]$$

$$\overline{\Psi}'(x) = (2\pi)^{-3/2} \sum_{s=\uparrow,\downarrow} \int d^3p \, (m_e/E)^{1/2} \cdot$$
$$\sum_{m=-\infty}^{\infty} b_m |2m+1|^3 \cdot \left[ b^\dagger(|2m+1|\mathbf{p},s) \cdot \overline{u}(\mathbf{p},s) \cdot e^{+i|2m+1|px} + d(|2m+1|\mathbf{p},s) \cdot \overline{v}(\mathbf{p},s) \cdot e^{-i|2m+1|px} \right]$$

A similar substitution can be made for photons:

$$\left. \begin{array}{l} a((2n+1)\mathbf{k},\lambda)\cdot\varepsilon_\mu\cdot e^{-i(2n+1)\,kx} \quad \Rightarrow \quad a^\dagger(|2n+1|\mathbf{k},\lambda)\cdot\varepsilon_\mu^*\cdot e^{+i|2n+1|\,kx} \\ a^\dagger((2n+1)\mathbf{k},\lambda)\cdot\varepsilon_\mu^*\cdot e^{+i(2n+1)\,kx} \quad \Rightarrow \quad a(|2n+1|\mathbf{k},\lambda)\cdot\varepsilon_\mu\cdot e^{-i|2n+1|\,kx} \end{array} \right\} \text{ for } n<0$$

This leads to a definition of the field operators $A'_\mu(x)$, where the factors $(2n+1)$ are replaced by $|2n+1|$:

$$(7')\quad A'_\mu(x) = (2\pi)^{-3/2} \sum_{\lambda=1,2} \int d^3k\, (2E)^{-\frac{1}{2}} \cdot$$
$$\cdot \sum_{n=-\infty}^{\infty} a_n |2n+1|^{5/2} \cdot \left[ a(|2n+1|\mathbf{k},\lambda)\cdot\varepsilon_\mu(\mathbf{k},\lambda)\cdot e^{-i|2n+1|\,kx} + a^\dagger(|2n+1|\mathbf{k},\lambda)\cdot\varepsilon_\mu^*(\mathbf{k},\lambda)\cdot e^{+i|2n+1|\,kx} \right]$$

In standard perturbation theory with sinusoidal waves, the substitutions made in (7′),(8′) lead to equivalent results for transition amplitudes and cross sections (via crossing symmetry). But this is not obvious when viewing anharmonic waves as composite particles. In the representation (8′) one might expect the charge of the electron to be reduced from the unit charge by converting negative harmonics into positrons. This problem is addressed in Appendix B by calculating the eigenvalue of the particle number operator. The lepton number actually remains 1, since one has a Fourier series not only for anharmonic single-electron states, but also for anharmonic operators. Along a similar vein, the mixing of particles and antiparticles in the representation (8′) seems unusual at a first glance. But a similar mixing of electron and hole wave functions occurs in the BCS theory of superconductivity during the Bogoliubov transformation. Thus it is safe to represent negative harmonics either way, using negative energy electrons in (7),(8) or positive energy positrons in (7′),(8′).

## 2. Feynman Rules for Anharmonic Quantum Fields

This new method needs to be tested against the well-confirmed results of standard quantum electrodynamics. Therefore it is useful to generalize the Feynman rules to anharmonic waves, even though it seems counterproductive to develop rules based on perturbation theory for non-perturbative method. The Feynman rules can be derived systematically from vacuum expectation values of time-ordered products of field operators [4]. This procedure gets quite involved if done thoroughly. Rather than reiterating the complete derivation, we will emphasize the new features that are specific to anharmonic waves (for an outline of the key aspects see Appendix C).

First of all, the standard Feynman diagrams need to be generalized to composite particles. An electron consists of a series of harmonics, which act like heavy electrons and positrons. The task is illustrated in Fig. 1A for a pair of composite electrons interacting with each other via a photon (dotted line). The two electrons are decomposed into their harmonics m and m′.

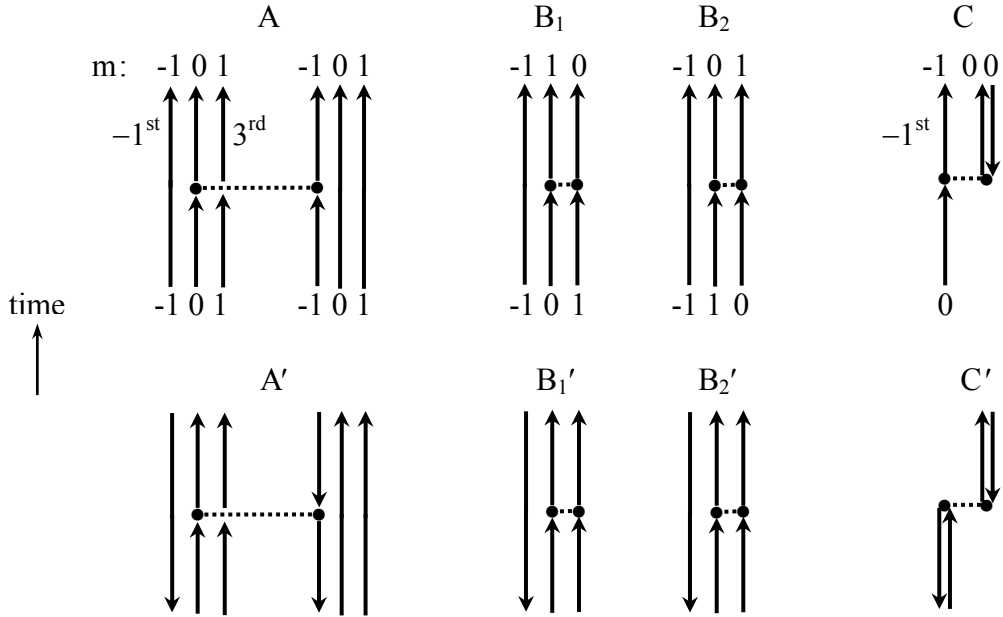

**Figure 1** Feynman diagrams for anharmonic waves, decomposed into harmonics. m=0 represents the fundamental, m=−1 the −1$^{st}$ harmonic, and m=+1 the 3$^{rd}$ harmonic.
A) External scattering between constituents of different particles. B) Internal scattering between constituents of the same particle. C) Internal generation of the −1$^{st}$ harmonic, together with an electron-positron pair. B and C are restricted to collinear momenta. The bottom row shows the alternative representations from (7′), where negative harmonics become positrons (with the arrows rotated by 180$^0$).

In the simplest type of diagram, the constituents of two particles interact pairwise with each other, as shown in Fig. 1A. This is similar to the quark picture of proton-proton scattering at high momentum transfer, where complex interactions between composite protons can be simplified to pairwise interactions between quarks. All combinations need to be taken into account, although the dominant contribution comes from the interaction between the two fundamentals (not shown). It is similar to the corresponding Feynman diagram in standard quantum electrodynamics, except that the Fourier coefficients $a_0, b_0$ are smaller than 1 due to the sum rule (5a). The second largest contribution comes from interactions between a fundamental and a lowest harmonic (m=±1), as shown in Fig. 1A.

In the alternative representation (7′),(8′) the arrows are rotated by $180^0$ for negative harmonics (Fig. 1A′).

The diagrams in Fig. 1 $B_1, B_2$ describe interactions between constituents of the same particle. Splicing $B_1$ and its inverse $B_2$ together leads to an oscillation between the two harmonics m = 0,1. This effect may be responsible for the energy and momentum oscillations that occur when the standard energy and momentum operators are applied to an anharmonic wave (see [3], Section 5).

It is also possible to construct internal diagrams involving only a single constituent, as in Fig. 1 C. This diagram generates the $-1^{st}$ harmonic together with an electron-positron pair and thus is related to the Fourier coefficients $b_{-1}$.

With the general structure of Feynman diagrams in hand we now focus on the elements of a Feynman diagram, i.e., external lines, internal lines, and vertices.

An external photon line consists of the polarization vector $\varepsilon_\mu(\mathbf{k},\lambda)$ according to the standard Feynman rules. With anharmonic waves one obtains a series of such diagrams for all the harmonics, multiplied by their Fourier coefficients $a_n$ (Fig. 2). The polarization vector remains unchanged, since all harmonics have the same polarization.

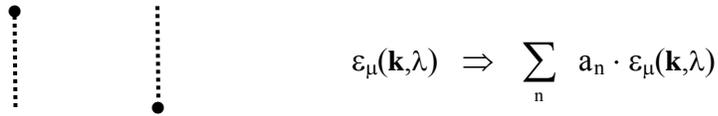

**Figure 2**   Generalized Feynman rule for external photons, either incoming (left) or outgoing (right). A single line becomes an infinite Fourier series of lines for the harmonics, each multiplied by a Fourier coefficient $a_n$.

External electron and positron lines are illustrated in Fig 3. For anharmonic waves one has again a series of external lines for all the harmonics, multiplied by their Fourier coefficients $b_m$. Each positive harmonic is represented by the spinor $u(\mathbf{p},s)$ or $v(\mathbf{p},s)$ of the fundamental, and each negative harmonic by the spinor $\hat{u}(\mathbf{p},s)$ or $\hat{v}(\mathbf{p},s)$. Although the momentum $\mathbf{p}$ is multiplied by (2m+1) for the harmonics, this factor cancels out between the numerators and denominators of the spinors. The representation (8′) converts $\hat{u}$ to v and $\hat{v}$ to u, as shown in the caption of Fig. 3. The corresponding arrows are rotated by $180^0$ about the vertex. The momentum $\mathbf{p}$ of the spinors $\hat{u}(\mathbf{p},s)$ and $\hat{v}(\mathbf{p},s)$ is not inverted, since they are defined in (A2′) such that $\mathbf{p}$ is the momentum of the fundamental (opposite to the momentum eigenvalue of a negative energy solution).

|  |  | $\sum_{m \geq 0} b_m \cdot u(\mathbf{p},s)$ |  |  | $\sum_{m \geq 0} b_m \cdot \bar{u}(\mathbf{p},s)$ |
|---|---|---|---|---|---|
| e⁻ in | u(**p**,s) ⇒ | $+\sum_{m<0} b_m \cdot \hat{u}(\mathbf{p},s)$ | e⁻ out | $\bar{u}(\mathbf{p},s)$ ⇒ | $+\sum_{m<0} b_m \cdot \hat{\bar{u}}(\mathbf{p},s)$ |
| e⁺ out | v(**p**,s) ⇒ | $\sum_{m \geq 0} b_m \cdot v(\mathbf{p},s)$ $+\sum_{m<0} b_m \cdot \hat{v}(\mathbf{p},s)$ | e⁺ in | $\bar{v}(\mathbf{p},s)$ ⇒ | $\sum_{m \geq 0} b_m \cdot \bar{v}(\mathbf{p},s)$ $+\sum_{m<0} b_m \cdot \hat{\bar{v}}(\mathbf{p},s)$ |

**Figure 3** Generalized Feynman rules for external electrons and positrons. A single line becomes an infinite series of lines for the harmonics, represented by the spinors u,v and the Fourier coefficients $b_m$. Negative harmonics require spinors for negative energies $(\hat{u},\hat{v})$. In the alternative representation (8′) they are converted to antiparticles with opposite four-momentum and spin by rotating the arrows $180^0$ about a vertex and substituting $\hat{u}(\mathbf{p},s) \Rightarrow v(\mathbf{p},-s)$, $\hat{v}(\mathbf{p},s) \Rightarrow u(\mathbf{p},-s)$, $\hat{\bar{u}}(\mathbf{p},s) \Rightarrow \bar{v}(\mathbf{p},-s)$, $\hat{\bar{v}}(\mathbf{p},s) \Rightarrow \bar{u}(\mathbf{p},-s)$.

Internal lines are associated with $g_{\mu\nu} \cdot i\tilde{D}(k)$, the Fourier transform of the photon propagator. In real space the photon propagator is defined as the vacuum expectation value of a time-ordered product of two field operators:

(9) $\quad iD_{\mu\nu}(x'-x) = \langle 0| T[A_\mu(x') A_\nu(x)] |0\rangle = \begin{cases} \langle 0| A_\mu(x') A_\nu(x) |0\rangle & \text{for } x'_0 > x_0 \\ \langle 0| A_\nu(x) A_\mu(x') |0\rangle & \text{for } x'_0 < x_0 \end{cases}$

The anharmonic propagator consists of the standard Feynman propagator, multiplied by the two Fourier coefficients $a_n \cdot a_{n'}$ of the two field operators:

(10) $\quad \tilde{D}_{\mu\nu}(k) = \int D_{\mu\nu}(y) \cdot e^{iky} d^4y = g_{\mu\nu} \cdot \tilde{D}(k) \qquad\qquad$ Feynman gauge

(11) $\quad \tilde{D}(k) = \sum_{n,n'=-\infty}^{\infty} a_n a_{n'} \cdot \dfrac{-4\pi}{k_\mu k^\mu + i\varepsilon} = \dfrac{-4\pi}{k_\mu k^\mu + i\varepsilon} \qquad\qquad \varepsilon \to 0^+$

The factor $4\pi$ in $\tilde{D}(k)$ originates from the Gaussian unit system (with $\alpha = e^2/\hbar c = e^2$ in units of $\hbar,c$). Due to the sum rule (5c) the sum over the double series becomes a factor 1. The end result is simply the photon propagator of standard quantum electrodynamics.

When the two harmonics n, n′ are different, the four-momentum changes from $(2n+1)\cdot k$ at one end of an internal line to $(2n'+1)\cdot k$ at the other. Consequently, the four-momentum is not conserved along an internal line. This mixing of harmonics by the propagator reflects the fact that they are not eigenstates of the momentum operator $i\partial/\partial x^\mu$ (compare [3], Section 5).

Internal electron lines are associated with $i\tilde{S}(p)$, the Fourier transform of the electron propagator. It is defined analogous to the photon propagator, but with a minus sign for $x'_0 < x_0$ due to the anti-commutators:

(12) $\quad i S_{\alpha\beta}(x'-x) = \langle 0| T[\Psi_\alpha(x') \overline{\Psi}_\beta(x)] |0\rangle = \begin{cases} +\langle 0| \Psi_\alpha(x') \overline{\Psi}_\beta(x) |0\rangle & \text{for } x'_0 > x_0 \\ -\langle 0| \overline{\Psi}_\beta(x) \Psi_\alpha(x') |0\rangle & \text{for } x'_0 < x_0 \end{cases}$

Once more, one obtains a double Fourier series multiplied by the standard electron propagator $\tilde{S}(p)$ of quantum electrodynamics. And the double Fourier series sums up again to a factor of 1 according to (5c):

(13) $\quad \tilde{S}(p) = \int S(y) \cdot e^{ipy} \, d^4y$

(14) $\quad \tilde{S}(p) = \sum_{m,m'=-\infty}^{\infty} b_m b_{m'} \cdot \dfrac{\gamma^\mu p_\mu + m_e}{p_\mu p^\mu - m_e + i\varepsilon} = \dfrac{\gamma^\mu p_\mu + m_e}{p_\mu p^\mu - m_e + i\varepsilon} \qquad \varepsilon \to 0^+$

The structure of the anharmonic electron propagator $\tilde{S}$ is resembles the propagator obtained for the anharmonic solution of an electron in the potential of an electromagnetic wave [5],[6]. It consists of a double series of Fourier coefficients, multiplied by the propagator of the fundamental (see [6], Equation (9)). In that case the potential of the electromagnetic wave is external, which keeps the Dirac equation linear.

The vertex is characterized by a factor $e$ and the Dirac matrices $\gamma^\mu$, which connect three quantum field operators from the adjacent electron and photon lines:

(15) $\quad -i e \cdot A_\mu(x) \cdot \overline{\Psi}(x) \gamma^\mu \Psi(x)$

The fields and their Fourier coefficients are already included in external and internal lines. That leaves the standard vertex $-ie\gamma^\mu$ and a δ-function for momentum conservation in the fundamental. All the harmonics produce the same condition (see Appendix C):

(16) $\quad p' - p = k$

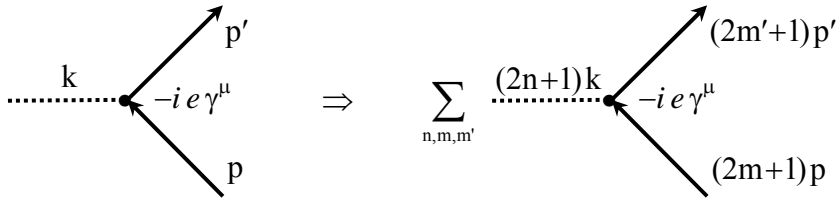

**Figure 4**  Vertex diagram for anharmonic waves. The four-momentum is conserved for the fundamental but generally violated for the harmonics.

Since the harmonics n,m,m′ connecting to a vertex can take all possible combinations, their four-momenta are generally not conserved (see Fig. 4). This reflects the non-linear interaction at a vertex, which scrambles the harmonics. A similar situation is encountered along internal lines. The propagator is always taken for the momentum of the fundamental, while the harmonics can take all possible combinations. Therefore, it is sufficient to draw Feynman diagrams only for the fundamental and just list the possible Fourier coefficients for the harmonics with the internal and external lines. This has been done in Fig. 5.

So far we have argued that anharmonic internal lines and vertices are described by the same Feynman rules as in standard quantum electrodynamics. External lines still carry Fourier coefficients, however, which seem to prevent complete agreement. Actually, it is possible to pair up external electron lines to form product series of the type (5c) that add up to 1. This can be seen in Fig. 5, where two external lines are always connected by a well-defined series of internal electron lines with aligned arrows. This holds, because each vertex contains a pair of aligned arrows and no other electron line.

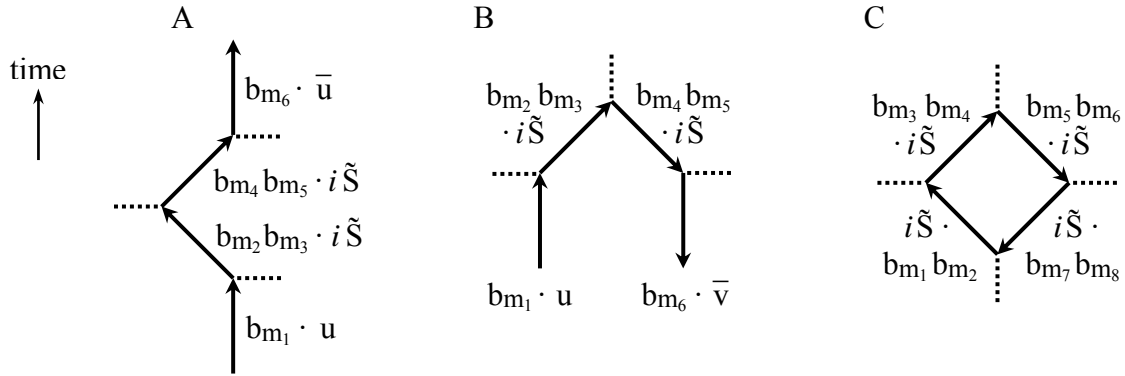

**Figure 5** Simplified Feynman diagrams for anharmonic waves, consisting of the diagram for the fundamental plus labels for the Fourier coefficients for the harmonics. Electron and positron lines form a continuous sequence of aligned arrows, while photon lines are isolated. External electron lines come in pairs. Several possible topologies are shown:
A) Line,  B) Half-loop,  C) Loop

External photon lines can be paired as well, since diagrams with an odd number of external photon lines vanish due to Furry's theorem [4]. Using again the sum rule (5c) for a double series, it becomes clear that the Feynman rules for anharmonic waves give the same result as the standard Feynman rules of quantum electrodynamics.

### 3. Feynman Diagrams for the Anharmonicity

For obtaining exact quantum field operators and single-particle wave functions it is necessary to find the Fourier coefficients of the anharmonic plane waves. Particularly important are the two lowest harmonics with n=±1 or m=±1. Higher harmonics can be expected to fall off like a power of the coupling constant $\alpha$. Already in standard quantum electrodynamics one can find Feynman diagrams that represent the generation of harmonics, as shown in Fig. 6 for photons and Fig. 7 for electrons.

The Feynman diagram for elastic photon-photon scattering in Fig. 6A has been known for quite some time as Delbrück scattering [7]. It can be rearranged to describe the generation of harmonics, as shown in Fig. 6B. In that case all three incoming photons have equal momenta.

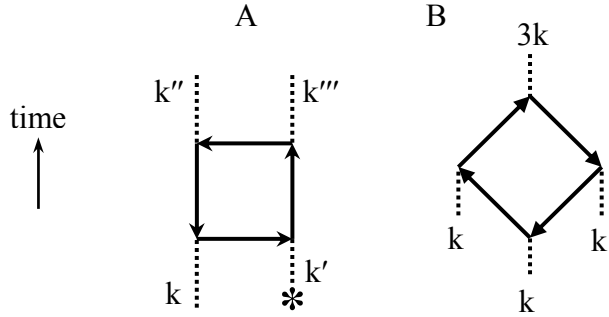

**Figure 6** Anharmonic processes for photons:
A) Splitting of a photon by the strong electric field of a nucleus (represented by a star). This process has been observed in [14].
B) 3$^{rd}$ harmonic generation from three photons.

The cross section for these diagrams is very low. They are of the order $\alpha^2$, and they require MeV photons to come close to the threshold for producing real electron-positron pairs. Laser photons have energies in the eV range, and radio waves are down to μeV. That makes the observation of anharmonic effects for photons in vacuum very difficult. As a consequence, photon-photon scattering itself has not yet been detected [7]-[13]. However, the related process of photon splitting in Fig. 6A has actually been observed [14]. A Gamma photon (k) can scatter off a virtual photon (k′) from a heavy nucleus (star) and generate two outgoing Gammas (k″,k‴).

Diagrams for the generation of the $-1^{st}$ and $3^{rd}$ harmonic of an electron are shown in Fig. 7. They contain only two vertices and thus are linear in $\alpha$. One can expect such diagrams to dominate over the photon diagrams in Fig. 6. An additional electron-positron pair is required, because lepton number and charge need to be conserved. The momenta need to be collinear to satisfy energy and momentum conservation for a standard

Feynman diagram. The leg with negative energy (–p) for the –1st harmonic in Fig. 7A can be rotated to become an incoming positron. The result is a diagram for electron-positron scattering in the forward direction (compare Fig. 1 C,C′).

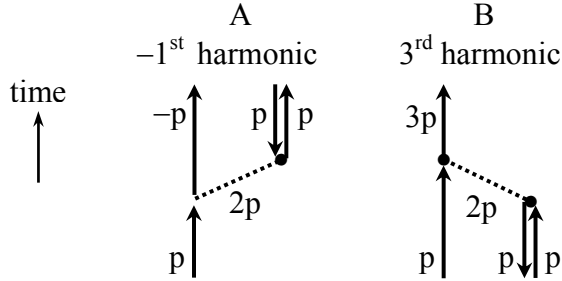

**Figure 7** Generation of harmonics for electrons:
A) The –1st harmonic plus an $e^+e^-$ pair created from the fundamental.
B) The 3rd harmonic created from an electron plus an $e^+e^-$ pair.

All harmonics of an electron higher than the –1st are off the mass shell, in contrast to the harmonics of a photon. They correspond to heavy electrons or positrons with mass $|2m+1| \cdot m_e$ and thus are not detectable in isolation. Indirect detection methods are required, where a harmonic is generated and annihilated in the same diagram. One can construct such diagrams by placing the diagrams in Fig. 7 back-to-back with their time-reversed versions. Even that is not a realistic situation for anharmonic waves. One would have to consider complete anharmonic waves interacting with each other. Otherwise the charge and lepton number would not be integer. This task is much more complicated than the diagrams discussed here.

## 4. Conclusions and Outlook

This work intends to flesh out the concept of using anharmonic plane waves as *ansatz* for interacting field operators in quantum field theory [3]. Although anharmonic waves are meant to be used in non-perturbative situations, it is wise to test them first in perturbation theory. Therefore, the Feynman rules of quantum electrodynamics are generalized to anharmonic fields. Each leg in a Feynman diagram becomes a series of harmonics which behave like heavy copies of a particle and propagate with the same phase velocity as the fundamental. For each combination of harmonics, the invariant amplitude is obtained by a straightforward extension of the standard Feynman rules. After summing over all harmonics the amplitude coincides with the result of standard quantum electrodynamics. This follows from a sum rule that originates from the condition $\psi^*\psi=1$, one of the three basic requirements for anharmonic plane waves [3].

This finding ensures that all the precise tests of quantum electrodynamics apply to anharmonic waves as well. The sum rule also keeps the lepton number and the charge integer, even though an electron is broken up into harmonics with non-integer Fourier coefficients.

A first attempt at obtaining the Fourier coefficients of anharmonic waves gives a qualitative answer. It is possible to find diagrams describing the generation of a harmonic in standard quantum electrodynamics. These are of $O(\alpha)$ for electrons and of $O(\alpha^2)$ for photons, suggesting that anharmonicity effects can be neglected for photons in lowest order. However, new concepts need to be developed to actually calculate the Fourier coefficients.

Since the coupling constant $\alpha$ is small, one can expect higher harmonics to be suppressed by powers of $\alpha$. For practical purposes, only a few harmonics should be able to provide insight into non-perturbative phenomena, such as high field situations and the question about the internal stability of an electron.

Once the Fourier coefficients are known, one has the exact field operators for a homogeneous electron system in hand. Inserting these into the Dyson-Schwinger equations produces algebraic equations for the coupling constant $\alpha$, the only remaining dimensionless parameter. If there is a non-trivial solution, the elusive fine structure constant $\alpha$ would become accessible to computation. To get started, one could insert a strongly-truncated anharmonic series into the Dyson-Schwinger equations or into various non-perturbative approximations, such as Hartree-Fock-Dirac [15],[16] or relativistic density functional theory [16],[17].

**Appendix A:   Anharmonic Quantum Field of a Dirac Spinor**

The definition of the anharmonic quantum field $\Psi(x)$ of a Dirac spinor proceeds analogous to the method used for a scalar field $\Phi(x)$ in [3]. The starting point is the conventional expansion of a spinor field $\Psi$ into plane waves:

(A1) $$\Psi(x) \;=\; (2\pi)^{-3/2} \int d^3p \sum_{s=\uparrow,\downarrow} (m_e/E)^{1/2} \cdot \left( b(\mathbf{p},s)\cdot u(\mathbf{p},s)\, e^{-i\,p\,x} \;+\; d^\dagger(\mathbf{p},s)\cdot v(\mathbf{p},s)\, e^{+i\,p\,x} \right)$$

$b$ = electron annihilation operator  $\qquad$ $b^\dagger$ = electron creation operator
$d$ = positron annihilation operator  $\qquad$ $d^\dagger$ = positron creation operator

$$\{b(\mathbf{p},s), b^\dagger(\mathbf{p}',s')\} = \{d(\mathbf{p},s), d^\dagger(\mathbf{p}',s')\} = \delta_{s,s'}\, \delta^3(\mathbf{p}-\mathbf{p}') \qquad s = \text{spin}$$

$$p^\mu = (p^0, \mathbf{p}) \qquad E = |p^0| = (\mathbf{p}^2 + m_e^2)^{1/2} \qquad p_\mu p^\mu = m_e^2 \qquad p\,x = p_\mu x^\mu$$

$u(\mathbf{p},s)$, $v(\mathbf{p},s)$ are the standard four-component Dirac spinors for electrons and positrons with positive energy ($p_0 > 0$). $v$ is obtained by charge conjugation of $u$: $v = u^C = i\gamma^2 u^*$

(A2) 
$$u(\mathbf{p},\uparrow) = [(E+m_e)/2m_e]^{1/2} \cdot \left(1, 0, \frac{p_z}{E+m_e}, \frac{p_x + i p_y}{E+m_e}\right) \qquad \text{positive energy}$$

$$u(\mathbf{p},\downarrow) = [(E+m_e)/2m_e]^{1/2} \cdot \left(0, 1, \frac{p_x - i p_y}{E+m_e}, \frac{-p_z}{E+m_e}\right)$$

$$v(\mathbf{p},\uparrow) = [(E+m_e)/2m_e]^{1/2} \cdot \left(\frac{p_x - i p_y}{E+m_e}, \frac{-p_z}{E+m_e}, 0, 1\right)$$

$$v(\mathbf{p},\downarrow) = [(E+m_e)/2m_e]^{1/2} \cdot \left(\frac{p_z}{E+m_e}, \frac{p_x + i p_y}{E+m_e}, 1, 0\right) \cdot (-1)$$

(A3) 
$$(i\gamma^\mu \partial_\mu - m_e) \cdot u(\mathbf{p},s)\, e^{-ipx} = 0 \qquad\qquad (\gamma^\mu p_\mu - m_e) \cdot u(\mathbf{p},s) = 0$$

$$(i\gamma^\mu \partial_\mu - m_e) \cdot v(\mathbf{p},s)\, e^{+ipx} = 0 \qquad\qquad (\gamma^\mu p_\mu + m_e) \cdot v(\mathbf{p},s) = 0$$

(A4) 
$$\bar{u}(\mathbf{p},s) \cdot u(\mathbf{p},s') = +\delta_{s,s'} \qquad\qquad \bar{u}(\mathbf{p},s) \cdot v(\mathbf{p},s') = 0$$

$$\bar{v}(\mathbf{p},s) \cdot v(\mathbf{p},s') = -\delta_{s,s'} \qquad\qquad \bar{v}(\mathbf{p},s) \cdot u(\mathbf{p},s') = 0$$

(A5) 
$$\sum_{s=\uparrow,\downarrow} u(\mathbf{p},s) \otimes \bar{u}(\mathbf{p},s) = \frac{\gamma^\mu p_\mu + m_e}{2m_e}$$

$$\sum_{s=\uparrow,\downarrow} v(\mathbf{p},s) \otimes \bar{v}(\mathbf{p},s) = \frac{\gamma^\mu p_\mu - m_e}{2m_e}$$

(A6) 
$$\bar{u}(\mathbf{p},s)\, \gamma^\mu\, u(\mathbf{p},s') = p^\mu/m_e \cdot \delta_{s,s'} \qquad\qquad \bar{u}(-\mathbf{p},s)\, \gamma^0\, v(\mathbf{p},s') = 0$$

$$\bar{v}(\mathbf{p},s)\, \gamma^\mu\, v(\mathbf{p},s') = p^\mu/m_e \cdot \delta_{s,s'} \qquad\qquad \bar{v}(-\mathbf{p},s)\, \gamma^0\, u(\mathbf{p},s') = 0$$

For negative harmonics one needs plane waves with inverted four-momentum, since $e^{-ipx}$ is interchanged with $e^{+ipx}$. These are negative energy states with $-p_\mu$ as eigenvalue of $i\partial_\mu$. In order to have a common label $p_\mu$ for both negative and positive harmonics, we use the four-momentum $p^\mu = (p^0, \mathbf{p})$ of the fundamental as label for both positive and negative energy states. The wave functions for negative energy electrons $\hat{u}(\mathbf{p},s) \cdot e^{+ipx}$ satisfy the Dirac equation (A3'), and their energy eigenvalue $-p_0$ is negative. The spinors $\hat{v}$ for positrons with negative energy are obtained by charge conjugation of $\hat{u}$, i.e. $\hat{v} = \hat{u}^C = i\gamma^2 \hat{u}^*$. The wave functions $\hat{v}(\mathbf{p},s) \cdot e^{-ipx}$ satisfy the Dirac equation as well.

(A2′) $\quad \hat{u}(\mathbf{p},\uparrow) = [(E+m_e)/2m_e]^{1/2} \cdot \left( \dfrac{p_z}{E+m_e}, \dfrac{p_x+ip_y}{E+m_e}, 1, 0 \right) \quad$ negative energy

$\qquad \hat{u}(\mathbf{p},\downarrow) = [(E+m_e)/2m_e]^{1/2} \cdot \left( \dfrac{p_x-ip_y}{E+m_e}, \dfrac{-p_z}{E+m_e}, 0, 1 \right)$

$\qquad \hat{v}(\mathbf{p},\uparrow) = [(E+m_e)/2m_e]^{1/2} \cdot \left( 0, 1, \dfrac{p_x-ip_y}{E+m_e}, \dfrac{-p_z}{E+m_e} \right) \cdot (-1)$

$\qquad \hat{v}(\mathbf{p},\downarrow) = [(E+m_e)/2m_e]^{1/2} \cdot \left( 1, 0, \dfrac{p_z}{E+m_e}, \dfrac{p_x+ip_y}{E+m_e} \right)$

(A3′) $\quad (i\gamma^\mu \partial_\mu - m_e) \cdot \hat{u}(\mathbf{p},s)\, e^{+ipx} = 0 \qquad\qquad (\gamma^\mu p_\mu + m_e) \cdot \hat{u}(\mathbf{p},s) = 0$

$\qquad (i\gamma^\mu \partial_\mu - m_e) \cdot \hat{v}(\mathbf{p},s)\, e^{-ipx} = 0 \qquad\qquad (\gamma^\mu p_\mu - m_e) \cdot \hat{v}(\mathbf{p},s) = 0$

(A4′) $\quad \bar{\hat{u}}(\mathbf{p},s) \cdot \hat{u}(\mathbf{p},s') = -\delta_{s,s'} \qquad\qquad \bar{\hat{u}}(\mathbf{p},s) \cdot \hat{v}(\mathbf{p},s') = 0$

$\qquad \bar{\hat{v}}(\mathbf{p},s) \cdot \hat{v}(\mathbf{p},s') = +\delta_{s,s'} \qquad\qquad \bar{\hat{v}}(\mathbf{p},s) \cdot \hat{u}(\mathbf{p},s') = 0$

(A5′) $\quad \displaystyle\sum_{s=\uparrow,\downarrow} \hat{u}(\mathbf{p},s) \otimes \bar{\hat{u}}(\mathbf{p},s) = \dfrac{\gamma^\mu p_\mu - m_e}{2m_e}$

$\qquad \displaystyle\sum_{s=\uparrow,\downarrow} \hat{v}(\mathbf{p},s) \otimes \bar{\hat{v}}(\mathbf{p},s) = \dfrac{\gamma^\mu p_\mu + m_e}{2m_e}$

(A6′) $\quad \bar{\hat{u}}(\mathbf{p},s)\, \gamma^\mu\, \hat{u}(\mathbf{p},s') = p^\mu/m_e \cdot \delta_{s,s'} \qquad \bar{\hat{u}}(\mathbf{p},s)\, \gamma^0\, \hat{v}(-\mathbf{p},s') = 0$

$\qquad \bar{\hat{v}}(\mathbf{p},s)\, \gamma^\mu\, \hat{v}(\mathbf{p},s') = p^\mu/m_e \cdot \delta_{s,s'} \qquad \bar{\hat{v}}(-\mathbf{p},s)\, \gamma^0\, \hat{u}(\mathbf{p},s') = 0$

For mixed products between positive and negative energy spinors one obtains the following results with the phase factor chosen in (A2′):

(A4″) $\quad \bar{u}(\mathbf{p},\uparrow) \cdot \hat{v}(\mathbf{p},\downarrow) = \bar{v}(\mathbf{p},\downarrow) \cdot \hat{u}(\mathbf{p},\uparrow) = \bar{\hat{u}}(\mathbf{p},\uparrow) \cdot v(\mathbf{p},\downarrow) = \bar{\hat{v}}(\mathbf{p},\downarrow) \cdot u(\mathbf{p},\uparrow) = +1$

$\qquad \bar{u}(\mathbf{p},\downarrow) \cdot \hat{v}(\mathbf{p},\uparrow) = \bar{v}(\mathbf{p},\uparrow) \cdot \hat{u}(\mathbf{p},\downarrow) = \bar{\hat{u}}(\mathbf{p},\downarrow) \cdot v(\mathbf{p},\uparrow) = \bar{\hat{v}}(\mathbf{p},\uparrow) \cdot u(\mathbf{p},\downarrow) = -1$

All other combinations vanish.

(A5″) $\quad \displaystyle\sum_{s=\uparrow,\downarrow} \hat{u}(\mathbf{p},s) \otimes \bar{u}(\mathbf{p},s) + \hat{v}(\mathbf{p},s) \otimes \bar{v}(\mathbf{p},s) = +\gamma^5$

$\qquad \displaystyle\sum_{s=\uparrow,\downarrow} u(\mathbf{p},s) \otimes \bar{\hat{u}}(\mathbf{p},s) + v(\mathbf{p},s) \otimes \bar{\hat{v}}(\mathbf{p},s) = -\gamma^5$

(A6″) $\bar{u}(\mathbf{p},\uparrow)\gamma^\mu \hat{v}(\mathbf{p},\downarrow) = \bar{v}(\mathbf{p},\uparrow)\gamma^\mu \hat{u}(\mathbf{p},\downarrow) = \bar{\hat{u}}(\mathbf{p},\downarrow)\gamma^\mu v(\mathbf{p},\uparrow) = \bar{\hat{v}}(\mathbf{p},\downarrow)\gamma^\mu u(\mathbf{p},\uparrow) = + p^\mu/m_e$

$\bar{u}(\mathbf{p},\downarrow)\gamma^\mu \hat{v}(\mathbf{p},\uparrow) = \bar{v}(\mathbf{p},\downarrow)\gamma^\mu \hat{u}(\mathbf{p},\uparrow) = \bar{\hat{u}}(\mathbf{p},\uparrow)\gamma^\mu v(\mathbf{p},\downarrow) = \bar{\hat{v}}(\mathbf{p},\uparrow)\gamma^\mu u(\mathbf{p},\downarrow) = - p^\mu/m_e$

$\bar{u}(-\mathbf{p},s)\gamma^0 \hat{u}(\mathbf{p},s') = \bar{v}(-\mathbf{p},s)\gamma^0 \hat{v}(\mathbf{p},s') = \bar{\hat{u}}(-\mathbf{p},s)\gamma^0 u(\mathbf{p},s') = \bar{\hat{v}}(-\mathbf{p},s)\gamma^0 v(\mathbf{p},s') = 0$

$\bar{u}(-\mathbf{p},s)\gamma^0 \hat{u}(\mathbf{p},s') = \bar{v}(-\mathbf{p},s)\gamma^0 \hat{v}(\mathbf{p},s') = \bar{\hat{u}}(-\mathbf{p},s)\gamma^0 u(\mathbf{p},s') = \bar{\hat{v}}(-\mathbf{p},s)\gamma^0 v(\mathbf{p},s') = 0$

The plane wave expansion of the standard Dirac field (A1) is generalized by replacing sinusoidal waves with anharmonic waves and expanding those into the Fourier series (4) of Class 2 sinusoidal waves, as discussed in [3], Section 7:

(A7) $\boxed{e^{i\,p\,x} \;\Rightarrow\; \sum_{m=-\infty}^{\infty} b_m \cdot e^{i(2m+1)\,p\,x}}$ for Class 2 anharmonic waves

The harmonics require the following substitutions:

(A8) $p_\mu \;\Rightarrow\; (2m+1)\cdot p_\mu \qquad E \Rightarrow |2m+1|\cdot E \qquad m_e \Rightarrow |2m+1|\cdot m_e$

$\int d^3 p \;\Rightarrow\; \int d^3((2m+1)\cdot p) = \int d^3 p \cdot |2m+1|^3$

$(m_e/E)^{-1/2} \;\Rightarrow\; (|2m+1|m_e/|2m+1|E)^{-1/2} = (m_e/E)^{-1/2}$

$b(\mathbf{p},s) \;\Rightarrow\; b((2m+1)\mathbf{p},s)$

$d(\mathbf{p},s) \;\Rightarrow\; d((2m+1)\mathbf{p},s)$

$u(\mathbf{p},s) \;\Rightarrow\; u((2m+1)\mathbf{p},s) = u(\mathbf{p},s) \qquad$ for $m \geq 0$

$v(\mathbf{p},s) \;\Rightarrow\; v((2m+1)\mathbf{p},s) = v(\mathbf{p},s)$

$u(\mathbf{p},s) \;\Rightarrow\; \hat{u}(|2m+1|\mathbf{p},s) = \hat{u}(\mathbf{p},s) \qquad$ for $m < 0$

$v(\mathbf{p},s) \;\Rightarrow\; \hat{v}(|2m+1|\mathbf{p},s) = \hat{v}(\mathbf{p},s)$

$p = (p^0, \mathbf{p})$ always refers to the four-momentum of the fundamental. Note the distinction between the Fourier coefficients $b_m$ and the annihilation operators $b((2m+1)\mathbf{p},s)$. Compared to scalar and vector fields, there is an extra factor $m_e^{1/2}$ in (A1). It takes care of the different dimension of spinors compared to vectors and scalars. The ratio $(m_e/E)^{-1/2}$ is not affected by the factor $|2m+1|$. The momentum integration $\int d^3((2m+1)\cdot p)$ generates a factor $|2m+1|^3$ when converted to $\int d^3(p)$. In the Fourier series one needs to use $u, v, b, d$ for positive harmonics (with positive energy) and $\hat{u}, \hat{v}, \hat{b}, \hat{d}$ for negative harmonics (with negative energy). The factor $|2m+1|$ appears in both the numerators and the denominators of the spinors and cancels out (see their definitions (A2),(A2′)). This leads to the expansion of an anharmonic Dirac field operator given in (8):

(A9) $\Psi(x) = (2\pi)^{-3/2} \sum_{s=\uparrow,\downarrow} \int d^3p \, (m_e/E)^{1/2} \cdot$

$\cdot \begin{bmatrix} \sum_{m\geq 0} b_m |2m+1|^3 \cdot \left[ b((2m+1)\mathbf{p},s) \cdot u(\mathbf{p},s) \cdot e^{-i(2m+1)px} + d^{\dagger}((2m+1)\mathbf{p},s) \cdot v(\mathbf{p},s) \cdot e^{+i(2m+1)px} \right] \\ + \sum_{m<0} b_m |2m+1|^3 \cdot \left[ \hat{b}((2m+1)\mathbf{p},s) \cdot \hat{u}(\mathbf{p},s) \cdot e^{-i(2m+1)px} + \hat{d}^{\dagger}((2m+1)\mathbf{p},s) \cdot \hat{v}(\mathbf{p},s) \cdot e^{+i(2m+1)px} \right] \end{bmatrix}$

The conjugate operator $\overline{\Psi}$ is expanded into conjugated operators, spinors, and plane waves. As a result of the harmonics, an electron becomes a composite particle consisting of the bare electron with mass $m_e$ accompanied by heavy copies with mass $|2m+1| \cdot m_e$.

**Appendix B: Negative Harmonics and the Particle Number Operator**

This appendix provides a quantitative answer to the questions raise at the end of Section 1. Is there a problem with fractional lepton numbers? Does the alternative representation of the field operators in (8′) change the lepton number? These questions are addressed by calculating the eigenvalues of the particle number operator for single particle states. Thereby we follow the outline given for a real scalar field in [3], Section 7, Equations (24)-(31).

The standard lepton number operator of an electron with momentum $\mathbf{p}$ and spin s is given by:

(B1) $\quad N = b^{\dagger}(\mathbf{p},s) \, b(\mathbf{p},s) - d^{\dagger}(\mathbf{p},s) \, d(\mathbf{p},s)$

For anharmonic waves one needs to generalize the creation and annihilation operators. This is achieved by recalling that the anharmonic creation operator $b^{\dagger}_a(\mathbf{p},s)$ for an electron generates the state $|\psi_a(\mathbf{p},s)\rangle$ which is described by the anharmonic wave function $\psi_a(\mathbf{p},s)$. Such a state consists of a Fourier series of normalized sinusoidal plane wave states $|\mathbf{p},s\rangle$:

(B2) $\quad \psi_a(\mathbf{p},s) = (2\pi)^{-3/2} \sum_{m\geq 0} b_m \cdot u(\mathbf{p},s) \cdot e^{-i(2m+1)px} + \sum_{m<0} b_m \cdot \hat{u}(\mathbf{p},s) \cdot e^{-i(2m+1)px}$

(B3) $\quad |\psi_a(\mathbf{p},s)\rangle = \sum_{m\geq 0} b_m |(2m+1)\mathbf{p},s\rangle + \sum_{m<0} b_m |(2m+1)\mathbf{p},s\rangle_n = b^{\dagger}_a(\mathbf{p},s) |0\rangle$

$|0\rangle$ is the vacuum state. The right side defines an anharmonic creation operator. Therefore, it consists a Fourier series of creation operators for sinusoidal states:

(B4) $\quad b^{\dagger}_a(\mathbf{p},s) = \sum_{m\geq 0} b_m \cdot b^{\dagger}((2m+1)\mathbf{p},s) + \sum_{m<0} b_m \cdot \hat{b}^{\dagger}((2m+1)\mathbf{p},s)$

This holds for the anharmonic field operators developed in Appendix A and listed in (8). The alternative version (8′) produces a different expansion, because the negative harmonics are represented by positron solutions:

(B4′) $\quad b^{\dagger}_a{}'(\mathbf{p},s) = \sum_{m\geq 0} b_m \cdot b^{\dagger}(|2m+1|\mathbf{p},s) \ + \ \sum_{m<0} b_m \cdot d(|2m+1|\mathbf{p},-s)$

Analogous expansions exist for the other operators $d^{\dagger}_a(\mathbf{p},s)$, $d_a(\mathbf{p},s)$, and $b_a(\mathbf{p},s)$.

In order to obtain the anharmonic lepton number operator $N_a$, these expansions are inserted into the generalization of (B1) to anharmonic waves:

(B5) $\quad N_a = b^{\dagger}_a(\mathbf{p},s)\, b_a(\mathbf{p},s) \ - \ d^{\dagger}_a(\mathbf{p},s)\, d_a(\mathbf{p},s)$

With the expansions (B3) and (B4),(B4′) for the anharmonic states and operators one can determine the eigenvalue of $N_a$ for the state $|\psi_a(\mathbf{p},s)\rangle$ that contains a single anharmonic electron. First we consider the version (B4) which contains negative energy states:

(B6) $\quad N_a |\psi_a(\mathbf{p},s)\rangle \ = \ (\ b^{\dagger}_a(\mathbf{p},s)\, b_a(\mathbf{p},s) - d^{\dagger}_a(\mathbf{p},s)\, d_a(\mathbf{p},s)\ )\ |\psi_a(\mathbf{p},s)\rangle$

The positron annihilation operator $d_a(\mathbf{p},s)$ on the right side generates the number 0, since the Fourier series for $|\psi_a(\mathbf{p},s)\rangle$ does not contain any positron states. The electron annihilation operator $b_a(\mathbf{p},s)$ on the left converts $|\psi_a(\mathbf{p},s)\rangle$ to the vacuum state $|0\rangle$. This is worked out explicitly for all four combinations of positive and negative harmonics:

$b_a(\mathbf{p},s)\, |\psi_a(\mathbf{p},s)\rangle =$

$= \sum_{m'\geq 0, m''\geq 0} b_{m'} b_{m''} \cdot b((2m'+1)\mathbf{p},s)\, |(2m''+1)\mathbf{p},s\rangle + \sum_{m'<0, m''<0} b_{m'} b_{m''} \cdot \hat{b}((2m'+1)\mathbf{p},s)\, |(2m''+1)\mathbf{p},s\rangle$

$+ \sum_{m'\geq 0, m''<0} b_{m'} b_{m''} \cdot b((2m'+1)\mathbf{p},s)\, |(2m''+1)\mathbf{p},s\rangle + \sum_{m'<0, m''\geq 0} b_{m'} b_{m''} \cdot \hat{b}((2m'+1)\mathbf{p},s)\, |(2m''+1)\mathbf{p},s\rangle$

$= \sum_{m'\geq 0, m''\geq 0} b_{m'} b_{m''} \cdot \delta_{m',m''}\, |0\rangle \ + \ \sum_{m'<0, m''<0} b_{m'} b_{m''} \cdot \delta_{m',m''}\, |0\rangle$

$= \sum_{m'=-\infty}^{\infty} b_{m'}{}^2\, |0\rangle \ = \ |0\rangle$

The two combinations of harmonics with equal sign in the first line have non-vanishing diagonal elements. The other two combinations in the second line do not contain annihilation operators matched with the corresponding harmonics and thus vanish. The sum over the diagonal elements gives 1 due to the sum rule (5a). One gets $|\psi_a(\mathbf{p},s)\rangle$ back from the vacuum state $|0\rangle$ by applying the anharmonic electron creation operator $b^{\dagger}_a(\mathbf{p},s)$ defined in (B3):

(B7) $\quad N_a |\psi_a(\mathbf{p},s)\rangle \ = \ b^{\dagger}_a(\mathbf{p},s)\, |0\rangle \ = \ 1 \cdot |\psi_a(\mathbf{p},s)\rangle$

As a result, the eigenvalue of the lepton number operator $N_a$ remains 1 for anharmonic waves, as it should.

The same calculation can be performed for the alternative representation (B4′) where negative harmonics are represented by positron states of the form $||2m+1|\mathbf{p},-s\rangle_p$:

(B6′) $\quad N'_a |\psi'_a(\mathbf{p},s)\rangle \;=\; \left(\, b'^\dagger_a(\mathbf{p},s)\, b'_a(\mathbf{p},s) - d'^\dagger_a(\mathbf{p},s)\, d'_a(\mathbf{p},s)\,\right)\, |\psi'_a(\mathbf{p},s)\rangle$

The action of the anharmonic annihilation operators on both electron and positron states is considered explicitly, because positrons are involved in the expansion of $|\psi'_a(\mathbf{p},s)\rangle$:

$b'_a(\mathbf{p},s)\, |\psi'_a(\mathbf{p},s)\rangle =$

$= \sum_{m'\geq 0, m''\geq 0} b_{m'} b_{m''} \cdot b(|2m'+1|\mathbf{p},s)\, ||2m''+1|\mathbf{p},s\rangle \;+\; \sum_{m'<0, m''<0} b_{m'} b_{m''} \cdot d(|2m'+1|\mathbf{p},-s)\, ||2m''+1|\mathbf{p},-s\rangle_p$

$+ \sum_{m'\geq 0, m''<0} b_{m'} b_{m''} \cdot b(|2m'+1|\mathbf{p},s)\, ||2m''+1|\mathbf{p},-s\rangle_p \;+\; \sum_{m'<0, m''\geq 0} b_{m'} b_{m''} \cdot d(|2m'+1|\mathbf{p},-s)\, ||2m''+1|\mathbf{p},s\rangle$

$= \sum_{m'\geq 0, m''\geq 0} b_{m'} b_{m''} \cdot \delta_{m',m''} |0\rangle \;+\; \sum_{m'<0, m''<0} b_{m'} b_{m''} \cdot \delta_{m',m''} |0\rangle$

$= \sum_{m'=-\infty}^{\infty} b_{m'}^{\,2}\, |0\rangle \;=\; 1 \cdot |0\rangle$

The first sum is the same as in the previous case. The second sum pairs positron annihilation operators with positron states, which gives the same result as pairing electron annihilation operators with electrons previously. The mixed sums in the second line vanish since either electron annihilation operators act on positron states or positron annihilation operators act on electron states. The action of the anharmonic positron annihilation operator takes the form:

$d'_a(\mathbf{p},s)\, |\psi'_a(\mathbf{p},s)\rangle =$

$= \sum_{m'\geq 0, m''\geq 0} b_{m'} b_{m''} \cdot d(|2m'+1|\mathbf{p},-s)\, ||2m''+1|\mathbf{p},s\rangle \;+\; \sum_{m'<0, m''<0} b_{m'} b_{m''} \cdot b(|2m'+1|\mathbf{p},s)\, ||2m''+1|\mathbf{p},-s\rangle_p$

$+ \sum_{m'\geq 0, m''<0} b_{m'} b_{m''} \cdot d(|2m'+1|\mathbf{p},-s)\, ||2m''+1|\mathbf{p},-s\rangle_p \;+\; \sum_{m'<0, m''\geq 0} b_{m'} b_{m''} \cdot b(|2m'+1|\mathbf{p},s)\, ||2m''+1|\mathbf{p},s\rangle$

$= \sum_{m'\geq 0, m''<0} b_{m'} b_{m''} \cdot \delta_{m',m''} |0\rangle \;+\; \sum_{m'<0, m''\geq 0} b_{m'} b_{m''} \cdot \delta_{m',m''} |0\rangle \;=\; 0$

Now the sums in the first line vanish due to a mismatch between operators and states, and the sums in the second line match electron operators with electrons and positron operators with positrons. However, the sums in the second line exclude $m'=m''$ and therefore vanish. Applying the creation operator $b'^\dagger_a(\mathbf{p},s)$ to the vacuum state gives:

(B7′) $\quad N'_a |\psi'_a(\mathbf{p},s)\rangle \;=\; b'^\dagger_a(\mathbf{p},s)\, |0\rangle \;=\; 1 \cdot |\psi'_a(\mathbf{p},s)\rangle$

Both definitions of the anharmonic field operators keep the lepton number integer.

**Appendix C: Generalizing the Feynman Rules to Anharmonic Fields**

Rather than going through the lengthy derivation of the standard Feynman rules, we will emphasize the features that are different for anharmonic waves. Those are used in Section 2 to generalize the Feynman rules for anharmonic waves. The following expressions should be viewed as a sketch of a much more involved procedure.

The Feynman rules are derived from the vacuum expectation value of a time-ordered product of quantum field operators [4]:

(C1) $\iint d^4x\, d^4x'... \langle 0| T[A_\mu(x) \cdot \overline{\Psi}(x)\, \gamma^\mu\, \Psi(x) \times A_\mu(x') \cdot \overline{\Psi}(x')\, \gamma^\mu\, \Psi(x') \times ...] |0\rangle$

The field operators are grouped into groups of three at a vertex and are integrated over the space-time coordinate x of the vertex. To generalize this scheme to anharmonic field operators $A_\mu$, $\overline{\Psi}$, $\Psi$ one simply has to insert their expansions into anharmonic plane waves. Using (7′),(8′) for the field operators and selecting a single vertex x and a specific combination of harmonics n, m, m′ produces an integral of the form:

(C2) $\int d^4x\, \int d^3(|2n+1|\cdot k)\, \int d^3(|2m'+1|\cdot p')\, \int d^3(|2m+1|\cdot p)\, e^{-i|2n+1|\, kx} \cdot e^{+i|2m'+1|\, p'x} \cdot e^{-i|2m+1|\, px}$

$\cdot \langle 0| T[a_\mu(|2n+1|\cdot k, x) \cdot \overline{\psi}(|2m'+1|\cdot p', x)\, \gamma^\mu\, \psi(|2m+1|\cdot p, x) |0\rangle$

Factors in the expansion of the field operators that are not directly relevant to the following arguments have been lumped into the terms $a_\mu$, $\overline{\psi}$, $\psi$. Likewise, the Fourier coefficients $a_n, b_m, b_{m'}$ have been omitted for clarity.

As a first step, the three exponential factors are combined under the four-dimensional integral over x, in order to obtain the δ-function for momentum conservation around a vertex. In a second step, the three momentum integrations are converted from the harmonics to the fundamental by changing the integration variables:

(C3) $|2n+1|\cdot \mathbf{k} \Rightarrow \mathbf{k}$ $\qquad |2m+1|\cdot \mathbf{p} \Rightarrow \mathbf{p}$ $\qquad |2m'+1|\cdot \mathbf{p'} \Rightarrow \mathbf{p'}$

This gives an integral without any vestiges of the harmonics:

(C4) $\iiint d^3k\, d^3p\, d^3p'\, \langle 0| T[a_\mu(k,x) \cdot \overline{\psi}(p',x)\, \gamma^\mu\, \psi(p,x) |0\rangle \cdot \int d^4x\, e^{-i(k+p'-p)x}$

$= \iiint d^3k\, d^3p\, d^3p'\, \langle 0| T[a_\mu(k,x) \cdot \overline{\psi}(p',x)\, \gamma^\mu\, \psi(p,x) |0\rangle \cdot (2\pi)^4 \cdot \delta^4(k+p'-p)$

$\Rightarrow\ p'-p = k$

Having reduced the momentum variables to those in standard quantum electrodynamics one can perform the integration over $d^4x$ to obtain a δ-function for energy and momentum conservation which is identical to that in standard quantum electrodynamics. Each harmonic generates the same δ-function as the fundamental. The only remnant of the harmonics is the product of their Fourier coefficients. These have been omitted here for clarity, but they are included in the rules for internal and external lines given in Section 2.